\documentclass[journal]{IEEEtran}
\usepackage{amsmath}
\usepackage{amssymb}
\usepackage{cite}
\usepackage{graphicx}
\usepackage[dvipsnames,table,xcdraw]{xcolor}
\usepackage{colortbl}
\usepackage{caption}
\usepackage{subcaption}
\usepackage{amsthm}
\usepackage{soul}
\usepackage{hhline}
\usepackage[labelformat=simple]{subcaption}
\usepackage{multirow}

\allowdisplaybreaks

\usepackage[latin1]{inputenc}           

\begin{document}

\title{A Complex-Valued Feedback Linearization-Based Controller for a Voltage Source Inverter Tied to the Grid via a Second-Order Filter}

\author{Gerardo Tapia-Otaegui$\dag$, Jorge~A.~Solsona$\ddag$,~\IEEEmembership{Senior~Member,~IEEE}, Sebastian Gomez Jorge$\ddag$, Ana~Susperregui$\dag$,~\IEEEmembership{Member,~IEEE}, Claudio A.~Busada$\ddag$, and M. Itsaso Martínez$\dag$\\
\thanks{This work was supported in part by the Spanish Ministry of Science and Innovation (project code PID2020-115484RA-I00), FEDER Funds, EU, in part by the Basque Government under research grant IT1644-22, in part by the Universidad Nacional del Sur (UNS), and in part by Consejo Nacional de Investigaciones Cient\'ificas y T\'ecnicas (CONICET).
$\dag$ Authors are with the Department of Automatic Control and Systems Engineering, University of the Basque Country UPV/EHU, Faculty of Engineering--Gipuzkoa, Donostia 20018, Spain.
$\ddag$ Authors are with the Instituto de Investigaciones en Ingenier\'ia El\'ectrica (IIIE), UNS-CONICET and Departamento de Ingenier\'ia El\'ectrica y de Computadoras, UNS, Bah\'ia Blanca, Argentina.
(e-mail: gerardo.tapia@ehu.eus; jsolsona@uns.edu.ar; sebastian.gomezjorge@uns.edu.ar; ana.susperregui@ehu.eus; cbusada@uns.edu.ar; mirenitsaso.martinez@ehu.eus).}
}

\maketitle

\begin{abstract}
In this document, a nonlinear control law for a grid-tied converter is introduced. The converter topology consists of a voltage source inverter (VSI) linked to the grid through an inductive-capacitive second-order filter, its input being connected to a capacitive DC-link supplied by a renewable energy-based input power source. In order to achieve good performance in presence of large state excursions caused mainly by substantial set-point modifications and/or considerable disturbances, a nonlinear control law based on a complex-valued feedback linearization strategy is designed. Specifically, a flat output is adopted, which is given by the summation of the energy stored in the DC-link capacitor and in the output filter's inductor and capacitor, as well as by the reactive energy at the output. After linearizing the system through a pertinent coordinate transformation and a nonlinear feedback, a linear trajectory tracking control law is implemented. The performance of the system controlled by applying the proposed strategy is tested via simulation for a very weak grid of unity \textit{X/R} ratio, yielding satisfactory results.

\end{abstract}


\section{Introduction}
Since the beginning of this century, the design criteria for grids managing electric energy have undergone a substantial change, giving priority to aspects that were previously considered secondary. To accompany this change, it has been necessary to develop new technologies that, for example, allow construction of electric generators that reduce greenhouse gas emissions; process electric power in such a way as to change the waveform, adapting amplitude and frequency levels in the transport and distribution networks; and optimizing consumption by adjusting the loads, including an electronic energy processing stage \cite{Fang2012}.

In order to execute the new functionalities of the grid, including those mentioned above, an essential device is the power electronic converter (PEC) \cite{Wang2024}. Power grids have been penetrated by this equipment, that is used in all the stages of the electric power management chain. The PEC is a constituent part of, among others, the following devices:
\begin{itemize}
	\item Generators that use primary sources based on non-conventional renewable energies ---inverter-based resources (IBRs) \cite{Haghighi2024}.
	
	\item HVDC transmission systems \cite{Zhang2017}.
	
	\item Solid-state transformers used in distribution \cite{She2013}.
	
	\item Electric drives \cite{Lim2024, Wang2020}.
	
	\item Equipment used to maintain the quality standards of the energy circulating in the grid ---for example, FACTS, STATCOMs, active power filters and digital voltage restorers, among others \cite{Chawda2020}.
	
\end{itemize}
Such is the degree of penetration of the PECs, that specialists have been forced to redefine and extend the definition and classification of the concepts of stability in power systems \cite{Hatziargyriou2021}.

As mentioned, PECs working in inverter mode are used to inject power into the grid from non-conventional renewable sources (i.e., IBRs). This is achieved by building a voltage or current source, which is coupled to the electrical grid. This type of source is made up of different components. Either a primary DC power source is connected to its input or a DC bus containing energy from one or more sources. Depending on the type of PEC, the DC bus can be inductive or capacitive. The intermediate stage consists of a set of semiconductor devices that switch the signal from the primary source, and constitute either a current- or a voltage-source converter \cite{Liu2020}. Finally, the output stage consists of an inductive filter, a capacitive one or a combination of both \cite{Loh2005}.

In IBRs, once the topology has been selected, it is necessary to establish a closed-loop control strategy to improve the behavior in the presence of disturbances and/or when it is desired to track time-varying references. The control algorithm can be classified into one of two main groups. Grid forming (GFM) \cite{Tayyebi2018, Tozak2024, Gajare2023, Song2022} refers to the control algorithm whose objective is to maintain a given voltage value at the point of common coupling (PCC), when the device works in voltage-source mode. On the other hand, grid feeding denotes the control algorithm used when the device injects power into the grid operating as a current source, which is synchronized or self-synchronized with the grid. There is a third type named grid supporting, where the device provides ancillary services. The so-called grid-following (GFL) mode includes the last two: grid feeding and grid supporting. 

There are several papers in the literature where different authors present control strategies for various types of topologies. In \cite{Pattabiraman2018}, a small-signal study is carried out to compare the behavior of an inverter controlled in GFL mode and one controlled in GFM mode, where the IBR is interconnected with a synchronous generator. It is possible to find several GFM algorithms. Among the most widespread, are those that emulate the behavior of other types of electrical or electronic equipment. Since a voltage generator that allows establishing a defined alternating voltage value in the PCC is sought, an algorithm that mimics the behavior of the synchronous generator described by the swing equation (synchronverter) has been proposed, as well as another that emulates the behavior of a nonlinear oscillator (virtual oscillator), for example reproducing the Van der Pol equation \cite{Lu2022, Lu_et_al2022}. Moreover, an algorithm called dispatchable virtual oscillator control (dVOC) has been employed, which emulates a linear oscillator, but also includes nonlinear terms that form an inner control-loop for current and another for voltage. In \cite{He2023}, the nonlinear stability problem of the GFM complex droop control is studied, where the dVOC algorithm is used to control the complex frequency at the PCC \cite{Ponce2024, Milano2022}. Several algorithms considering large-signal stability are presented and compared in \cite{Hosseinpour2023}, while a nonlinear control of the synchronverter algorithm, based on feedback linearization and backstepping, was presented in \cite{Lourenco2025}. Analyses of the transient stability in a grid containing several converters, when adopting both the synchronverter algorithm and the non-uniform Kuramoto oscillator, were presented in \cite{Dorfler2012}. In \cite{Smith2022}, the behaviors of three different types of controller are compared; namely, the dVOC, the droop control and the match controller.

On the other hand, a huge number of linear controllers whose design is based on cascaded control have been proposed. They generally consist of an inner current-loop and an outer voltage-loop, which is supposed to be slow with respect to the former one and fixes its current reference value \cite{Rosso2021}. There are also linear controllers that feed back the states of the output filter \cite{Tian2025, Dong2023, Zamani2023}. The authors of \cite{Mahmoud2022} present a review that compares this type of controllers. In general, linear controllers greatly limit the achievable closed-loop transient performance, which turns out to be poor in the presence of large excursions of the states ---due, for example, to the tracking of references that vary rapidly or to abrupt and large disturbances in the output voltage. For this reason, several designers have endeavored to develop nonlinear controllers allowing better transient performance \cite{Eskandari2020, She2024}. A controller based on nonlinear polytopic regions is found in \cite{Ojo2024}, whilst the authors propose a nonlinear power decoupling in \cite{Li2023}. Controllers designed through backstepping \cite{Karunaratne2024} and nonlinear predictive control \cite{Samanta2024} have also been proposed.

Regarding grid-tied inverters commanded via GFL control strategies, either grid feeding or grid supporting, many papers can be found in the literature ---see, for instance, \cite{Liu2020, GomezJorge2024, GomezJorge_arXiv} and the references therein. Among others, a previous work by the authors of this document has shown that a complex-valued feedback linearization strategy allows obtaining an excellent performance, even with an unusually small value of the DC-link capacitance, when the inverter is connected to the grid through a first-order ---inductive--- filter \cite{GomezJorge2024}. In addition, a similar controller considering connection to a weak grid can be found in \cite{GomezJorge_arXiv}, where the suggested solution succeeds in robustifying the controller by feeding back an estimate of the PCC voltage provided by a notch filter.

In this work, a nonlinear control strategy based on complex-valued flatness is introduced. The topology of the converter consists of an input stage that comes from a capacitive DC-link fed by a power source, a voltage-source inverter (VSI) and a second-order output LC filter, working in GFL mode. The design allows full input-output linearization, hence avoiding the appearance of zero dynamics, since a flat output is selected. This output is computed as the sum of the complex energy stored in the passive components, which is the time integral of the complex power balance; i.e., the time integral of the difference between the input active power feeding the DC link and the complex instantaneous power at the output capacitor.

The rest of this document is organized as follows. Starting from the three-phase model of the converter under study, Section \ref{sec:Model} derives its corresponding complex-valued model. Then, Section \ref{sec:Control} addresses both the design and tuning of the proposed nonlinear GFL control solution, whose validation in simulation is dealt with in Section \ref{sec:Simu} for a medium-voltage very weak grid. Lastly, a section of conclusions closes the document.

\section{Complex-Valued Model of the Converter} \label{sec:Model}
\begin{figure*}
	\centering
	\includegraphics[width=0.7\linewidth]{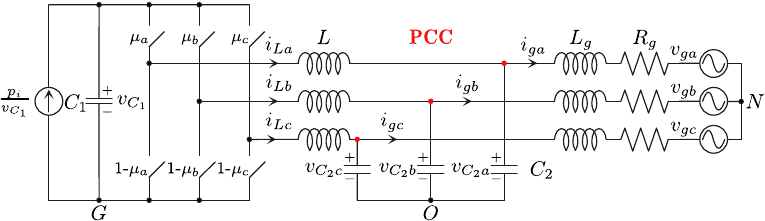}
	\caption{Schematic three-phase circuit of an inverter interfacing a RES with the grid through an LC filter.} \label{fig:Fig1esquema}
\end{figure*}

Figure \ref{fig:Fig1esquema} displays the schematic three-phase circuit of the system under consideration, which consists of an inverter interfacing a renewable energy source (RES) with the grid through an LC filter. The dynamic behavior corresponding to such conceptual model can therefore be described by the following set of state equations:
\begin{align}
	C_1\dot{v}_{C_1}  & = \frac{p_i}{v_{C_1}} - \left(\mu_ai_{La} + \mu_bi_{Lb} + \mu_ci_{Lc}\right) \label{eq:C1_vC1_dot} \\
	L\dot{i}_{La}     & = \mu_av_{C_1} - v_{C_2a} - v_{OG} \label{eq:L_iLa_dot} \\
	L\dot{i}_{Lb}     & = \mu_bv_{C_1} - v_{C_2b} - v_{OG} \label{eq:L_iLb_dot} \\
	L\dot{i}_{Lc}     & = \mu_cv_{C_1} - v_{C_2c} - v_{OG} \label{eq:L_iLc_dot} \\
	C_2\dot{v}_{C_2a} & = i_{La} - i_{ga} \label{eq:C2_vC2a_dot} \\
	C_2\dot{v}_{C_2b} & = i_{Lb} - i_{gb} \label{eq:C2_vC2b_dot} \\
	C_2\dot{v}_{C_2c} & = i_{Lc} - i_{gc} \label{eq:C2_vC2c_dot} \\
	L_g\dot{i}_{ga}   & = v_{C_2a} - R_gi_{ga} - v_{ga} - v_{NO} \label{eq:Lg_iga_dot} \\
	L_g\dot{i}_{gb}   & = v_{C_2b} - R_gi_{gb} - v_{gb} - v_{NO} \label{eq:Lg_igb_dot} \\
	L_g\dot{i}_{gc}   & = v_{C_2c} - R_gi_{gc} - v_{gc} - v_{NO}, \label{eq:Lg_igc_dot}
\end{align}
where $C_1$ is the DC-link capacitance, $L$ and $C_2$ are the inductance and capacitance of the LC filter, and $L_g$, $R_g$ and $v_{ga, b, c}$ correspond, respectively, to the inductance, resistance and three-phase voltage of the grid's Thévenin equivalent. Concerning the remaining three-phase variables, $\mu_{a, b, c}$ is the modulation index, $i_{La, b, c}$ and $v_{C_2a, b, c}$ are the inductor current and capacitor voltage of the LC filter, and $i_{ga, b, c}$ denotes the grid current. Finally, $v_{C_1}$ is the DC-link voltage, $p_i$ represents the power inputted from the RES ---modeled as a constant power source (CPS)---, while $v_{OG}$ and $v_{NO}$ refer to the voltage differences between points $O$ and $G$, and points $N$ and $O$ in Fig.\ \ref{fig:Fig1esquema}.

Moreover, Fig.\ \ref{fig:Fig1esquema} reveals that $i_{ga} + i_{gb} + i_{gc} = 0$ at all times, which, in turn, implies that $i_{La} + i_{Lb} + i_{Lc} = 0$ must also hold. Substitution of such restrictions in, respectively, (\ref{eq:Lg_iga_dot})--(\ref{eq:Lg_igc_dot}) and (\ref{eq:L_iLa_dot})--(\ref{eq:L_iLc_dot}) allows deriving $v_{NO}$ and $v_{OG}$ as given next:
\begin{align}
	v_{NO} \! & = \tfrac{1}{3}\!\left[v_{C_2a} + v_{C_2b} + v_{C_2c} - \left(v_{ga} + v_{gb} + v_{gc}\right)\right] \label{eq:v_NO} \\
	v_{OG} \! & = \tfrac{1}{3}\!\left[\left(\mu_a + \mu_b + \mu_c\right)\!v_{C1} - \left(v_{C_2a} + v_{C_2b} + v_{C_2c}\right)\right]\!. \label{eq:v_OG}
\end{align}

In order to reformulate (\ref{eq:C1_vC1_dot})--(\ref{eq:Lg_igc_dot}) in compact format, any three-phase $x_{a, b, c}$ variable is first expressed according to a stationary $\alpha$-$\beta$ reference frame by applying the power-invariant Clarke's transform, and then represented by a single complex variable, $\vec{x}$, as follows:
\begin{equation}
	\vec{x} = x_{\alpha} \! + \! jx_{\beta} = \sqrt{\frac{2}{3}}\!\left[x_a \! - \! \frac{x_b \! + \! x_c}{2} + j\frac{\sqrt{3}\left(x_b \! - \! x_c\right)}{2}\right]\!. \label{eq:x_complex_variable}
\end{equation}
Accordingly, application of (\ref{eq:x_complex_variable}) to (\ref{eq:C1_vC1_dot})--(\ref{eq:Lg_igc_dot}) leads to the following description of the grid-tied inverter dynamics:
\begin{align}
	C_1\dot{v}_{C1}        & = \frac{p_i}{v_{C1}} - \Re\!\left\{\vec{\mu}^{\ast}\vec{i}_L\right\} \label{eq:C1_vC1_dot_compact} \\
	L\dot{\vec{i}}_L       & = \vec{\mu}v_{C1} - \vec{v}_{C2} \label{eq:L_iL_dot} \\
	C_2\dot{\vec{v}}_{C2}  & = \vec{i}_L - \vec{i}_g \label{eq:C2_vC2_dot} \\
	L_g\dot{\vec{i}}_g     & = \vec{v}_{C2} - R_g\vec{i}_g - \vec{v}_{g}, \label{eq:Lg_ig_dot}
\end{align}
where $\vec{\mu} = \mu_{\alpha} + j\mu_{\beta}$, $\vec{i}_L = i_{L\alpha} + ji_{L\beta}$, $\vec{v}_{C2} = v_{C2\alpha} + jv_{C2\beta}$ and $\vec{v}_{g} = v_{g\alpha} + jv_{g\beta}$. On the other hand, $\Re\{\cdot\}$ and subscript $\ast$ refer, respectively, to the real part and the conjugate of the involved complex variable.

\section{Nonlinear Controller Design} \label{sec:Control}
The design of the control algorithm is approached by assuming that the grid's Thévenin equivalent is unknown, hence implying that state equation (\ref{eq:Lg_ig_dot}) cannot be used during the design process. Consequently, the controller design is based on the nonlinear sub-model given by the set of state equations (\ref{eq:C1_vC1_dot_compact})--(\ref{eq:C2_vC2_dot}).

\subsection{Feedback Linearization}
In this context, full input-output linearization of said nonlinear sub-model is achieved by selecting the following complex-valued energy-based flat output:
\begin{equation}
	\vec{\xi}_1 = \tfrac{1}{2}\!\left(C_1v_{C_1}^2 + L|\vec{i}_L|^2 + C_2|\vec{v}_{C_2}|^2\right) - j\!\textstyle\int_{0}^{t}q(\tau)d\tau, \label{eq:xi1}
\end{equation}
where $q = \Im\!\left\{\vec{v}_{C_2}\vec{i}_g^{\ast}\right\}$ is the reactive power injected into the grid at the PCC, with $\Im\{\cdot\}$ referring to the imaginary part of the involved complex variable. It should be stressed that the real part of $\vec{\xi}_1$ corresponds to the energy stored in the passive components of the sub-model; i.e., the DC-link capacitor, and both the inductor and capacitor of the LC filter.

Indeed, taking the time derivative of $\xi_1$ in (\ref{eq:xi1}) produces
\begin{equation}
	\dot{\vec{\xi}}_1 = C_1v_{C1}\dot{v}_{C_1} + L\Re\!\left\{\dot{\vec{i}}_L\vec{i}_L^{\ast}\right\} + C_2\Re\!\left\{\dot{\vec{v}}_{C_2}\vec{v}_{C_2}^{\,\ast}\right\} - jq. \label{eq:xi1_dot}
\end{equation}
Now, replacing (\ref{eq:C1_vC1_dot_compact})--(\ref{eq:C2_vC2_dot}) in (\ref{eq:xi1_dot}), it turns out that
\begin{equation}
	\dot{\vec{\xi}}_1 = p_i - \vec{v}_{C_2}\vec{i}_g^{\ast} = p_i - p - jq = \xi_2, \label{eq:xi2}
\end{equation}
where $p = \Re\!\left\{\vec{v}_{C_2}\vec{i}_g^{\ast}\right\}$ is the active power injected into the PCC. Similarly, computing the derivative of the first equality in (\ref{eq:xi2}) with respect to time and making use of (\ref{eq:C2_vC2_dot}) leads to
\begin{equation}
	\dot{\vec{\xi}}_2 = \dot{p}_i - \vec{v}_{C_2}\dot{\vec{i}}_g^{\ast} + \frac{\vec{i}_g - \vec{i}_L}{C_2}\vec{i}_g^{\ast} = \xi_3. \label{eq:xi3}
\end{equation}

Finally, taking the time derivative of (\ref{eq:xi3}) yields
\begin{equation}
	\dot{\vec{\xi}}_3 = \ddot{p}_i - \dot{\vec{v}}_{C_2}\dot{\vec{i}}_g^{\ast} - \vec{v}_{C_2}\ddot{\vec{i}}_g^{\ast} + \frac{\left(\dot{\vec{i}}_g - \dot{\vec{i}}_L\right)\!\vec{i}_g^{\ast} + \left(\vec{i}_g - \vec{i}_L\right)\!\dot{\vec{i}}_g^{\ast}}{C_2}, \label{eq:xi3_dot}
\end{equation}
and subsequent substitution of (\ref{eq:C2_vC2_dot}) and (\ref{eq:L_iL_dot}) in (\ref{eq:xi3_dot}) gives rise to the following expression:
\begin{equation}
	\dot{\vec{\xi}}_3 = \ddot{p}_i + \frac{2\!\left(\vec{i}_g - \vec{i}_L\right)}{C_2}\dot{\vec{i}}_g^{\ast} + \frac{\dot{\vec{i}}_g\vec{i}_g^{\ast}}{C_2} - \vec{v}_{C_2}\ddot{\vec{i}}_g^{\ast} - \frac{\vec{\mu}v_{C_1} - \vec{v}_{C_2}}{LC_2}\vec{i}_g^{\ast}, \label{eq:xi3_dot1}
\end{equation}
which is a nonlinear state equation explicitly depending on control input $\vec{\mu}$.

Consequently, as evidenced by (\ref{eq:xi2}), (\ref{eq:xi3}) and (\ref{eq:xi3_dot1}), selection of the $\vec{\xi}_1$ flat output in (\ref{eq:xi1}) allows attaining full input-output linearization of sub-model (\ref{eq:C1_vC1_dot_compact})--(\ref{eq:C2_vC2_dot}) as follows:
\begin{align}
	\dot{\vec{\xi}}_1 & = \vec{\xi}_2 \label{eq:xi1_dot1} \\
	\dot{\vec{\xi}}_2 & = \vec{\xi}_3 \label{eq:xi2_dot1} \\
	\dot{\vec{\xi}}_3 & = \vec{w}, \label{eq:xi3_dot2}
\end{align}
with auxiliary control input $\vec{w}$ being given by
\begin{equation}
	\vec{w} = \ddot{p}_i + \frac{2\!\left(\vec{i}_g - \vec{i}_L\right)}{C_2}\dot{\vec{i}}_g^{\ast} + \frac{\dot{\vec{i}}_g\vec{i}_g^{\ast}}{C_2} - \vec{v}_{C_2}\ddot{\vec{i}}_g^{\ast} - \frac{\vec{\mu}v_{C_1} - \vec{v}_{C_2}}{LC_2}\vec{i}_g^{\ast}. \label{eq:w}
\end{equation}

\subsection{State Feedback with Integral Action for Reference Tracking}
Based on the linear system expressed in companion form by (\ref{eq:xi1_dot1})--(\ref{eq:xi3_dot2}), the following state-feedback algorithm with integral action is proposed for auxiliary control input $\vec{w}$:
\begin{align}
	\vec{w} & = \dot{\vec{\xi}}_3^r - \vec{k}_3\vec{e}_{\xi_3} - \vec{k}_2\vec{e}_{\xi_2} - \vec{k}_1\vec{e}_{\xi_1} - \vec{k}_0\vec{y} \label{eq:w1} \\
	\vec{y} & = \textstyle\int_0^t \vec{e}_{\xi_1}\!(\tau)d\tau,
\end{align}
where superscript $r$ stands for reference value and, as a result, $\vec{e}_{\xi_1} = \vec{\xi}_1 - \vec{\xi}_1^r$, $\vec{e}_{\xi_2} = \vec{\xi}_2 - \vec{\xi}_2^r$ and $\vec{e}_{\xi_3} = \vec{\xi}_3 - \vec{\xi}_3^r$ represent the tracking errors in $\vec{\xi}_1$, $\vec{\xi}_2$ and $\vec{\xi}_3$, respectively.

As a first approximation, in order to establish $\vec{\xi}_1^r$, $\vec{\xi}_2^r$, $\vec{\xi}_3^r$ and $\dot{\vec{\xi}}_3^r$, it is assumed that the energy stored in both the inductor and capacitor of the LC filter is typically only a small fraction of that stored in the DC-link capacitor. Accordingly, starting from the reference values set for $v_{C_1}^r$ and $q^r$, as well as from their successive time derivatives, those reference values are approximated as follows:
\begin{align}
	\vec{\xi}_1^r       & = \tfrac{1}{2}C_1v_{C_1}^{r\, 2} - j\!\textstyle\int_{0}^{t}q^r(\tau)d\tau \label{eq:xi1_r} \\
	\vec{\xi}_2^r       & = \dot{\vec{\xi}}_1^r = C_1v_{C_1}^r\dot{v}_{C_1}^r - jq^r \label{eq:xi2_r} \\
	\vec{\xi}_3^r       & = \dot{\vec{\xi}}_2^r = C_1\!\left(\dot{v}_{C_1}^{r\, 2} + v_{C_1}^r\ddot{v}_{C_1}^r\right) - j\dot{q}^r \label{eq:xi3_r} \\
	\dot{\vec{\xi}}_3^r & = C_1\!\left(3\dot{v}_{C_1}^r\ddot{v}_{C_1}^r + v_{C_1}^r\!\dddot{v}_{\!\!C_1}^r\right) - j\ddot{q}^r. \label{eq:xi3_r_dot}
\end{align}
Specifically, it is desirable to define $\dot{v}_{C_1}^r$, $\ddot{v}_{C_1}^r$, $\dddot{v}_{\!\!C_1}^r$, $\dot{q}^r$ and $\ddot{q}_r$ so that, albeit rapidly, $v_{C_1}^r$ and $q^r$ evolve smoothly between different steady-state values. In particular, this work defines them in a way that such transitions follow first-order-like exponential transients.

On the other hand, focusing on $\vec{\xi}_1$, $\vec{\xi}_2$ and $\vec{\xi}_3$, which are also required to compute the auxiliary control input $\vec{w}$ in (\ref{eq:w1}), they should be derived by applying, respectively, (\ref{eq:xi1}), (\ref{eq:xi2}) and (\ref{eq:xi3}). Finally, the actual control input to be applied is derived from the auxiliary one by solving for $\vec{\mu}$ in (\ref{eq:w}), which leads to
\begin{equation}
	\vec{\mu} \! = \! \frac{LC_2\!\left(\ddot{p}_i \! - \! v_{C_2}\ddot{\vec{i}}_g^{\ast} \! - \! \vec{w}\right) \! + \! 2L\!\left(\vec{i}_g \! - \! \vec{i}_L\right)\!\dot{\vec{i}}_g^{\ast} \! + \! \left(\vec{v}_{C_2} \! + \! L\dot{\vec{i}}_g\right)\!\vec{i}_g^{\ast}}{v_{C_1}\vec{i}_g^{\ast}},
\end{equation}
where care must be taken to avoid division by zero if $v_{C_1}$ or/and $\vec{i}_g$ happen to be null.

It is worth pointing out that, in addition to the measurement of $v_{C_1}$, $\vec{i}_L$ and $\vec{v}_{C_2}$, knowledge of $p_i$, $\dot{p}_i$ and $\ddot{p}_i$, as well as that of $\vec{i}_g$, $\dot{\vec{i}}_g$ and $\ddot{\vec{i}}_g$, is required to implement the above-described control algorithm. In this context, the full-order disturbance observer proposed in \cite{Tapia2023} may, for instance, be adopted to estimate $p_i$, $\dot{p}_i$ and $\ddot{p}_i$. Analogously, $\vec{i}_g$, $\dot{\vec{i}}_g$ and $\ddot{\vec{i}}_g$ may also be estimated by means of either a full-order or a reduced-order observer.

Nonetheless, if it is assumed that $i_{ga, b, c}$ remains as a purely sinusoidal and balanced three-phase magnitude, it turns out that $\vec{i}_g = |\vec{i}_g|e^{j(\omega t + \varphi)}$, with $\omega$ representing the grid frequency. Accordingly, starting from the direct measurement of $\vec{i}_g$, its first and second time derivatives might be approximated by their respective steady-state values as
\begin{align}
	\dot{\vec{i}}_g  & = j\omega\vec{i}_g \label{eq:i_g_dot} \\
	\ddot{\vec{i}}_g & = -\omega^2\vec{i}_g. \label{eq:i_g_ddot}
\end{align}

\subsection{Tuning of State-Feedback Gains}
Substitution of (\ref{eq:w1}) in (\ref{eq:xi3_dot2}) leads to the following closed-loop tracking error dynamics:
\begin{equation}
	\dot{\vec{e}}_{\xi_3} + \vec{k}_3\vec{e}_{\xi_3} + \vec{k}_2\vec{e}_{\xi_2} + \vec{k}_1\vec{e}_{\xi_1} + \vec{k}_0\vec{y} = 0,
\end{equation}
which, by virtue of (\ref{eq:xi1_dot1}), (\ref{eq:xi2_dot1}) and the first equalities in both (\ref{eq:xi2_r}) and (\ref{eq:xi3_r}), can be expressed in the state-space matrix format given next:
\begin{equation}
	\begin{bmatrix}
		\dot{\vec{e}}_{\xi_1} \\
		\dot{\vec{e}}_{\xi_2} \\
		\dot{\vec{e}}_{\xi_3} \\
		\dot{\vec{y}}
	\end{bmatrix} =
	\underbrace{
		\begin{bmatrix}
			         0 &          1 &          0 &          0 \\
			         0 &          0 &          1 &          0 \\
			-\vec{k}_1 & -\vec{k}_2 & -\vec{k}_3 & -\vec{k}_0 \\
			         1 &          0 &          0 &          0
		\end{bmatrix}
	}_{\mathbf{A}}
	\begin{bmatrix}
		\vec{e}_{\xi_1} \\
		\vec{e}_{\xi_2} \\
		\vec{e}_{\xi_3} \\
		\vec{y}
	\end{bmatrix}. \label{eq:e_dot}
\end{equation}

Aiming to make tracking errors $\vec{e}_{\xi_1}$, $\vec{e}_{\xi_2}$ and $\vec{e}_{\xi_3}$, as well as $\vec{y}$, decay to zero following certain desired dynamics, the eigenvalues of matrix $\mathbf{A}$ can be assigned at one's discretion by suitably tuning state-feedback gains $\vec{k}_1$, $\vec{k}_2$, $\vec{k}_3$ and $\vec{k}_0$. To that end, MATLAB's `place' command may be applied, which calls for decomposing $\mathbf{A}$ as follows:
\begin{equation}
	\mathbf{A} = \mathbf{A}_1 - \mathbf{B}\mathbf{K},
\end{equation}
with
\begin{equation}
	\mathbf{A}_1 =
	\begin{bmatrix}
		0 & 1 & 0 & 0 \\
		0 & 0 & 1 & 0 \\
		0 & 0 & 0 & 0 \\
		1 & 0 & 0 & 0
	\end{bmatrix}\!,\
	\mathbf{B} =
	\begin{bmatrix}
		0 \\
		0 \\
		1 \\
		0
	\end{bmatrix}\!,\
	\mathbf{K}^T =
	\begin{bmatrix}
		\vec{k}_1 \\
		\vec{k}_2 \\
		\vec{k}_3 \\
		\vec{k}_0
	\end{bmatrix}
\end{equation}
and superscript $T$ denoting transpose. This way, the state-feedback gains in $\mathbf{K}$ can be adjusted by simply running the command $\mathbf{K} = \mathrm{place}(\mathbf{A}_1, \mathbf{B}, \mathbf{p})$, where $\mathbf{p}$ is a vector containing the target eigenvalues set for the $\mathbf{A}$ matrix.

\section{Simulation Results}  \label{sec:Simu}
\begin{figure}
	\centering
	\includegraphics[width=\columnwidth]{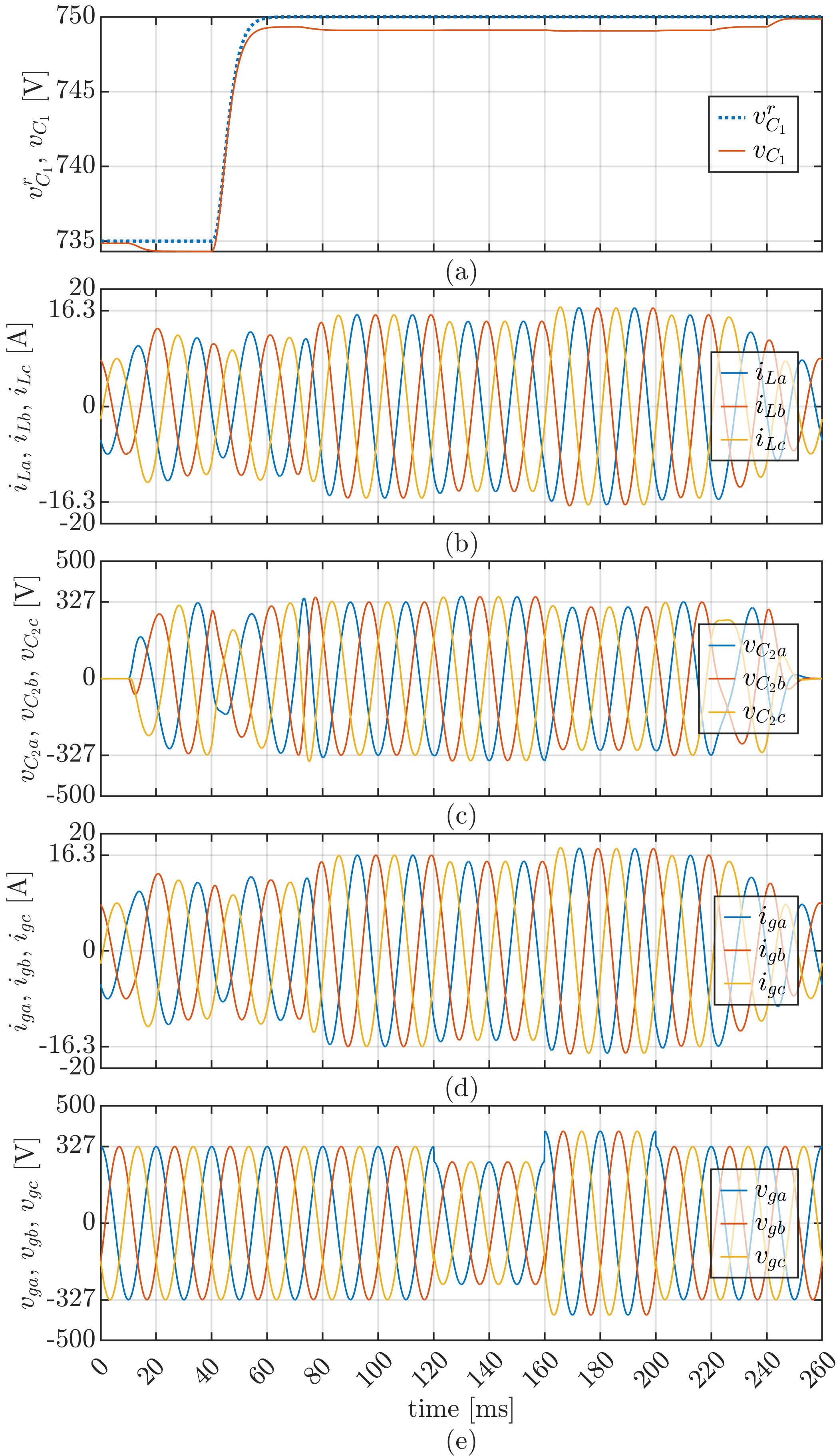}\\
	\caption{Simulation Results. (a) Reference and actual values of the DC-link voltage. (b) Three-phase current in the grid filter's inductor. (c) Three-phase voltage of the grid filter's capacitor. (d) Three-phase grid current. (e) Three-phase grid voltage.}\label{fig:sim1}
\end{figure}
\begin{figure}
	\centering
	\includegraphics[width=\columnwidth]{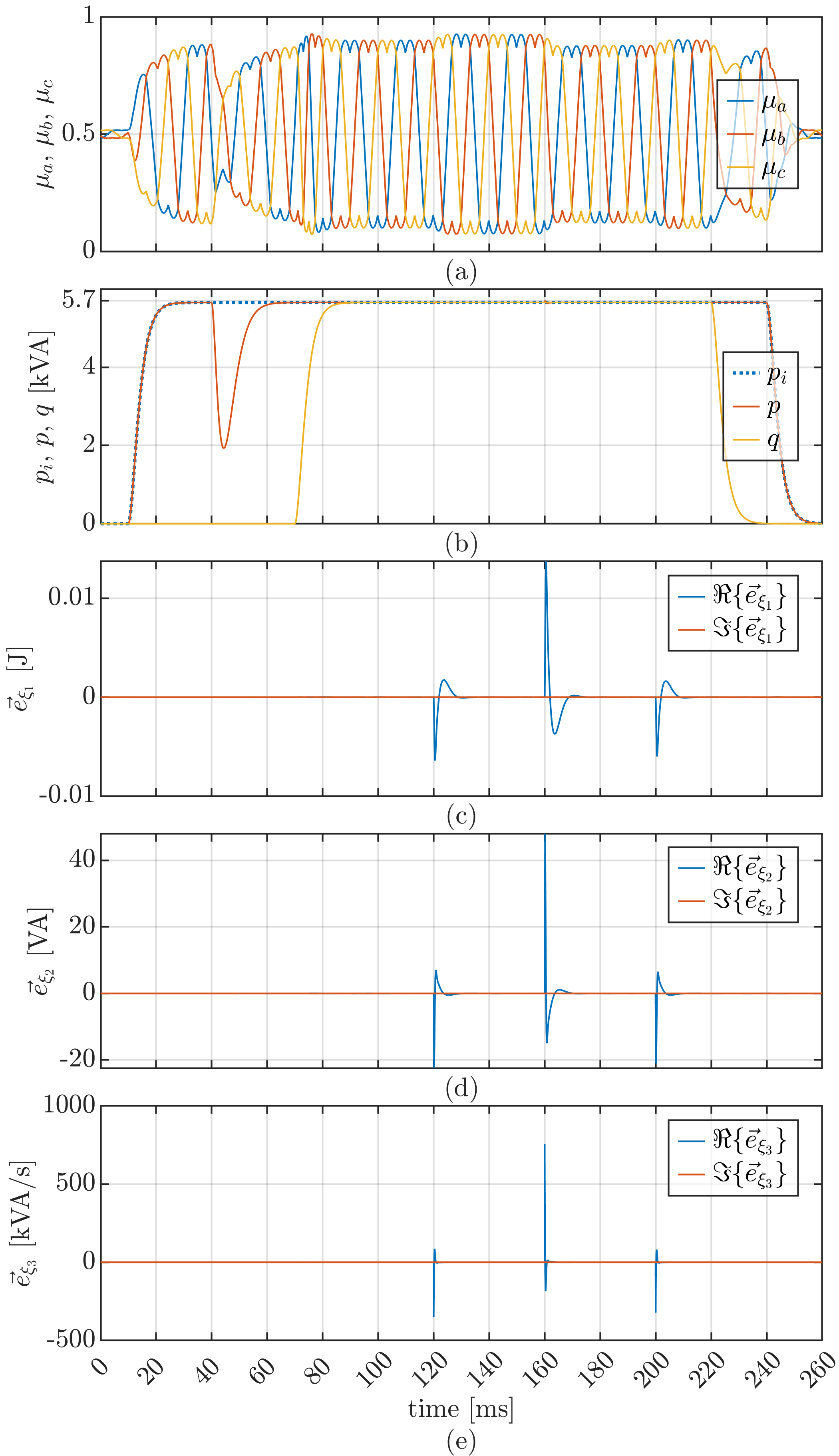}\\
	\caption{Simulation Results. (a) Three-phase modulation index. (b) Input power, and active and reactive powers injected into the PCC. (c) Real and imaginary parts of tracking error $\vec{e}_{\xi_1}$. (d) Real and imaginary parts of tracking error $\vec{e}_{\xi_2}$. (e) Real and imaginary parts of tracking error $\vec{e}_{\xi_3}$.}\label{fig:sim2}
\end{figure}
Simulation results are now presented for a converter whose parameters are as follows: $C_1 = 2.7\,$mF, $L = 5.7\,$mH, $C_2 = 9.9\,\mu$F, nominal apparent power $S_N = 8\,$kVA, nominal phase voltage of $230.94\,$Vrms and nominal angular frequency $\omega_N = 2\pi50\,$Hz. The grid impedance, given by $L_g = 90\,$mH and $R_g = 28.28\,\Omega$, corresponds to a very weak grid ---short-circuit ratio (SCR) $ = 0.5$--- of unity $X_g/R_g$ ratio, typical of medium voltage. The controller gains are tuned so that two pairs of complex conjugate poles are obtained in closed loop, with associated settling times of $1\,$ms and $10\,$ms, respectively, and damping coefficients of $0.707$ both. This results in $\vec{k}_1=4.28\times10^{10}$, $\vec{k}_2=5.12\times10^{7}$, $\vec{k}_1=1.01\times10^4$ and $\vec{k}_0=1.79\times10^{13}$.

The main results are displayed in Figs.\ \ref{fig:sim1} and \ref{fig:sim2}. As observable in Fig.\ \ref{fig:sim2}(b), the simulation test starts with $p_i = q = 0$. Under this condition, voltage $\vec{v}_{C_2} = 0$, and the grid current is dictated by both the grid voltage and the grid impedance. At $t=10\,$ms, the input power increases to $S_N/\sqrt2$ within $10\,$ms, as shown in Fig.\ \ref{fig:sim2}(b). As a result, the magnitudes of $\vec{v}_{C_2}$, $\vec{i}_L$ and $\vec{i}_g$ increase ---see Fig.\ \ref{fig:sim1}(b)--(d). The new magnitudes of $\vec{v}_{C_2}$ and $\vec{i}_L$ lead to an increase in the real part of $\vec{\xi}_1$. Since the energies stored in both $L$ and $C_2$ were neglected when setting $\vec{\xi}_1^r$, a small steady-state discrepancy arises between $v_{C_1}$ and $v_{C_1}^r$, as evidenced in Fig.\ \ref{fig:sim1}(a). Then, at $t = 20\,$ms, the DC-link voltage reference is increased from $735\,$V to $750\,$V within $10\,$ms. As a consequence, note that most of the input power is devoted to charge $C_1$, as reflected in Fig.\ \ref{fig:sim2}(b), where $p < p_i$ during the transient. Afterwards, at $t = 70\,$ms, the reactive power is increased to $S_N/\sqrt2$ within $10\,$ms, as observable in Fig.\ \ref{fig:sim2}(b), thus taking the total output apparent power to $S_N$. Consequently, both $\vec{i}_g$ and $\vec{v}_{C_2}$ increase to their respective nominal values of $16.33\,$Apeak and $326.6\,$Vpeak ---refer to Fig.\ \ref{fig:sim1}(c)-(d).

Now, aiming at evaluating the performance of the proposal to grid voltage variations, the magnitude of $\vec{v}_g$ is first decreased to $80\%$ of its nominal value at $t = 120\,$ms, then increased to $120\%$ of its nominal value at $t=160\,$ms, and finally brought back to its nominal value at $t=200\,$ms. Such step changes in the magnitude of $\vec{v}_g$, which are observable in Fig.\ \ref{fig:sim1}(e), lead to transients causing no significant variations in the system variables reflected in Fig.\ \ref{fig:sim1}(a)--(d) and Fig.\ \ref{fig:sim2}(b). Indeed, Fig.\ \ref{fig:sim2}(c)--(e) evidence that only some small tracking errors, $\vec{e}_{\xi_1}$--$\vec{e}_{\xi_3}$, arise in the transformed variables, which, in addition, extinguish rapidly. Later, the reactive and input powers are both decreased back to zero, at $t = 220\,$ms and $t = 240\,$ms, respectively, without any noteworthy effect on the main systems variables. Finally, Fig.\ \ref{fig:sim2}(a) displays the phase modulation indices, which do not saturate throughout the test.

\section{Conclusions}
A nonlinear control approach, based on complex-valued feedback linearization, has been described for GFL power converters interconnecting RESs with the electrical grid using LC filters. In particular, adoption of an instantaneous complex energy-dependent flat output makes full input-output linearization possible, which, in turn, allows achievement of outstanding closed-loop reference tracking performance by avoiding emergence of internal dynamics. Thereby, highly efficient transfer of active power from the RES to the PCC is accomplished, as well as high-dynamic-performance control of the reactive power exchanged at the PCC.

Complex formulation has been applied not only to describe the whole three-phase system to be governed, but also to design the state-feedback controller with integral action for reference tracking that is combined with the change of variables leading to full input-output linearization. The proposed control solution has been designed by assuming that the grid's Thévenin equivalent is unknown, its analytical tuning having been addressed by means of closed-loop eigenvalue assignment. Simulation tests carried out for a very weak grid of $0.5$ SCR and unity $X/R$ substantiate both the feasibility of the proposed control concept and the high-performance closed-loop dynamics to which it leads. 

In view of these promising simulation results, the authors intend to further deepen in the following aspects as future research:
\begin{itemize}
		\item It is planned to develop an observer in order to avoid both measuring the grid current and estimating its first and second time derivatives through the steady-state approximations in (\ref{eq:i_g_dot}) and (\ref{eq:i_g_ddot}). 
		\item Simulations reveal that the grid impedance must contain a minimum resistive fraction for the proposed control solution to be operational. A systematic stability analysis is already underway to determine the underlying cause(s) for the latter, as well as to adapt the proposal so that it retains its validity for (almost) purely inductive grids.
		\item Although the proposal exhibits high potential for very weak grids, it would be desirable to widen the range of SCRs for which it is effective.
		\item The strategy for setting instantaneous reference values for the flat output and its successive time derivatives should be extended for the case where the energy stored in the LC filter is not negligible with respect to that stored in the DC link.        
\end{itemize}

\bibliographystyle{IEEEtran}
\bibliography{biblioGFL}

\end{document}